\documentclass{iopart}

\usepackage{bm}
\usepackage{graphics}      

\expandafter\let\csname equation*\endcsname\relax 
\expandafter\let\csname endequation*\endcsname\relax 
\usepackage{amsmath}
\usepackage{epsfig}
\usepackage{graphics}
\usepackage{amsthm}
\usepackage{iopams}

\newcommand{\beq}{\begin{eqnarray}}
\newcommand{\eq}{\end{eqnarray}}

\newcommand{\lt}{\left}
\newcommand{\rt}{\right}
\newcommand{\n}{\nonumber}

\newcommand{\s}{\sigma}
\newcommand{\la}{\langle}
\newcommand{\ra}{\rangle}

\newcommand{\Dp}{\Delta^{(+)}}
\newcommand{\Dm}{\Delta^{(-)}}

\begin{document}

\date{\today}
\title{Diagrammatic 
approach to coherent backscattering of laser light by cold atoms:
Double scattering revisited}

\author{V. Shatokhin, T. Wellens and A. Buchleitner}
\address{Institute of Physics, University of Freiburg, Hermann-Herder-Str. 3,
D-79104 Freiburg, Germany}

\begin{abstract}
We present a derivation of the coherent backscattering
spectrum from two two-level atoms using the diagrammatic approach,
wherein the multiple scattering signal is deduced from single-atom
responses, and provide a physical interpretation of the single-atom
building blocks.
\end{abstract}
%



\section{Introduction}
Coherent backscattering (CBS) of light emerges due to the
constructive interference of multiply scattered counter-propagating
waves surviving the disorder average  \cite{sheng}. With optical
waves, CBS was successfully observed using classical -- {\em e.g.},
 polysterene
particles \cite{albada85} -- and quantum -- dilute cold atoms
\cite{labeyrie99} -- scatterers alike. A remarkable property of cold
atoms is their nonlinear {\it inelastic} scattering induced by a powerful resonant
laser field. Atomic saturation and inelastic scattering processes
accompanying it were shown to affect the CBS interference from cold
Sr (strontium) \cite{chaneliere04} and Rb (rubidium) \cite{balik05}
atoms, but a corresponding theory of CBS of intense laser light from
cold atomic clouds is still lacking.

The main challenge one has to deal with when describing CBS from cold
saturated atoms can be briefly summarized as follows.
A multiple scattering signal must be built on the basis of the accurately
described responses of individual scatterers to a strong laser field.
These responses can be found by solving the optical Bloch equations (OBE)
\cite{cohen-tannoudji}. However, a standard generalization of the OBE
to the $N$-atom case leads to a Lehmberg-type master equation governing
the evolution of the reduced density operator of all atoms which are laser-driven
and dipole-dipole interacting \cite{lehmberg70,agarwal_book}, with the number
of equations for the atomic averages growing exponentially with the number
of scatterers. So far, such a master equation in the context of CBS has
been solved only for $N=2$ atoms \cite{shatokhin05,shatokhin07}.

Recently, we have initiated an alternative method of generalizing
the OBE to the many-atom case which we call the diagrammatic approach to CBS \cite{geiger09,wellens10,geiger10}. In its
framework, the double scattering signal can be obtained from the
solutions of the OBE for an atom subjected to a bichromatic
classical driving. One component thereof represents the laser field
and another one the field scattered from the second atom.
Single-atom responses to a bichromatic field were evaluated
non-perturbatively in the laser field and perturbatively -- up to a
second order -- in the scattered field amplitude. Furthermore, by
self-consistently combining single-atom responses (to which we will
also refer as `building blocks'), we were able to derive analytical
expressions for the background and interference spectra of double
scattering. The spectra thus evaluated were rigorously shown to be
equivalent to that deduced on the basis of the two-atom master
equation \cite{shatokhin10}.

The motivation of the present contribution is twofold. First,
previously we focused on a
 discussion of the inelastic building blocks and their
 self-consistent combination into double scattering diagrams
 \cite{wellens10,geiger10}. These blocks determine the CBS signal
 only in the case of a very strong laser driving. In the present
 contribution
we will consider a general case and present a diagrammatic
derivation of the CBS spectra for arbitrary intensity of the laser
field. Second, the single-atom building blocks have not been given a
physical interpretation. Here, we
furnish the single atom building blocks with a physical
interpretation and establish a close connection with the results of
Mollow \cite{mollow72}.

The paper is organized as follows. In the next section we present
the building blocks contributing to the double scattering background and
interference spectra. Thereafter it is shown how these blocks can be
evaluated by solving the OBE under bichromatic driving. Thereby we
establish the connection to the method used by Mollow in
\cite{mollow72}. In Sec.~\ref{sec:connect} we formulate rules for
combining single-atom building blocks into double scattering
diagrams, present the full set of diagrams contributing to the elastic
and inelastic spectra, and give the explicit expressions thereof. We
conclude our work in Sec.~\ref{summary}.

\section{Single-atom building blocks}
\subsection{Graphical representation of the elastic and inelastic building blocks}
\label{sec:graph-single}
\begin{figure}
\includegraphics[width=7cm]{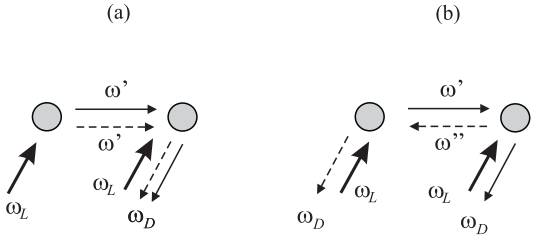}
\caption{Diagrammatic representation of the double scattering
processes contributing to the background (a) and interference (b)
spectra of CBS. The laser wave (thick arrows) of frequency $\omega_L$ is
scattered by the atoms (gray circles) into waves whose frequencies
$\omega'$, $\omega''$ and $\omega_D$ may differ from $\omega_L$.}
\label{fig:processes}
\end{figure}

To be self-contained, we will briefly outline here the main idea of
the diagrammatic approach to CBS of laser light from two two-level
atoms which was presented in detail in \cite{wellens10,geiger10}.
The fundamental double scattering processes surviving the disorder
averaging and contributing to the CBS background and interference
signals are shown in Fig.~\ref{fig:processes}.
We assume that the laser field may be sufficiently strong to
saturate the atomic transitions. Accordingly, the frequencies
$\omega'$, $\omega''$ and $\omega_D$ of the waves scattered by the
atoms towards each other and a detector can, but
need not, differ
from the laser frequency $\omega_L$, {\it i.e.}, correspond to inelastic
scattering. The co-propagating positive and negative frequency
amplitudes (solid and dashed arrows, respectively) contribute to the
background double scattering intensity, see Fig.~\ref{fig:processes}(a),
which is independent from the observation direction, whereas the
counter-propagating ones contribute to the CBS interference, see
Fig.~\ref{fig:processes}(b). We will be interested in finding the
frequency distributions of the background and interference spectra
(intensity vs. $\omega_D$) as a function of the driving field
parameters (such as the Rabi frequency and the offset from the atomic
transition frequency).

In the framework of the diagrammatic approach
\cite{geiger09,wellens10,geiger10}, these frequency distributions
can be derived on the basis of single-atom building blocks. The
latter represent spectral responses of an atom subjected to a {\it
classical} bichromatic driving field. The first component thereof
corresponds to a laser field of arbitrary strength, whereas a
second, weak, component describes the far-field scattered by
the other atom. Following the nomenclature used in laser 
spectroscopy \cite{cohen-tannoudji,xu2007}, we will refer to the laser and weak field
components as the pump and probe fields, respectively. While the classical description of the laser field
is common, applying the same description to the atomic radiation
exhibiting photon antibunching \cite{kimble77}
is, in general, wrong. However, in the dilute regime, when the double scattering
originates from exchange of a single photon, the validity of the
semiclassical treatment of the atom-probe field interaction does not
contradict the nonclassical character of the scattered field
\cite{wellens10,geiger10},
which manifests itself only in correlations between at least two photons.
Furthermore, the validity of the
classical ansatz for the probe field was proven analytically by
establishing the equivalence between the results of the diagrammatic
and master equation calculations \cite{shatokhin10}.

\begin{figure}
\includegraphics[width=6cm]{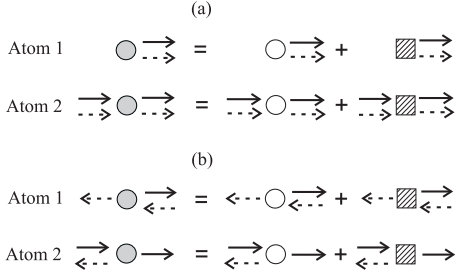}
\caption{Decomposition of scattering processes at each of the atoms
 (gray circles) from Fig.~\ref{fig:processes} as a sum of elastic
 (open circles) and inelastic (hatched squares) building blocks. Processes
 contributing to (a)  background, and (b) interference intensities. Frequencies
 of the probe and scattered fields will be defined after the expressions for the building blocks on the right hand side, as well as the rules for combining the building blocks are known.}
 \label{fig:decomp}
\end{figure}

The classical ansatz for the fields exchanged between the atoms
allows us to consider two atoms with their incoming and outgoing
fields separately from each other. The left hand sides in
Fig.~\ref{fig:decomp}(a) and (b) show the
decomposition
of the background
and interference contributions to CBS from
Fig.~\ref{fig:processes}(a) and (b), respectively, into single-atom
blocks. In order not to overburden diagrams, in
Fig.~\ref{fig:decomp} only fields scattered by the atoms are
depicted, but here and henceforth one should remember that the atoms
are also laser-driven and, furthermore, the effect of the laser field on atoms
is accounted
for non-perturbatively. Note that the arrows are not labeled by their
frequencies in Fig.~\ref{fig:decomp}. The latter will be defined in
the course of the subsequent analysis.

In general, for any finite value of the saturation parameter, there
are non-zero cross-sections for elastic and inelastic
scattering of photons
by a single laser-driven atom \cite{cohen-tannoudji}.
The same
also holds in presence of
additional probe field(s). In fact, one of the purposes of the
diagrammatic approach is to calculate elastic and inelastic responses
of a single laser-driven atom subject to probe fields.

Accordingly, the right hand sides of Fig.~\ref{fig:decomp} represent
decompositions of the total single-atom responses into elastic
(blank circles) and inelastic (hatched squares) building blocks.
Different shapes of the blocks
({\em i.e.}, circles or squares)
emphasize that the corresponding
expressions result from different equations of motion (for the
atomic dipole averages and temporal correlation functions, see
\ref{appA}) describing the elastic and inelastic scattering, respectively.
Different colours refer to photons emitted at the laser frequency (blank),
or at the frequency different from the laser frequency (hatched), as will be explained in Sec.~\ref{sec:graph-elastic}.

There is an important difference between the diagrams in
Figs.~\ref{fig:processes} and \ref{fig:decomp}. While the former
diagrams depict general (background and interference) double scattering processes for a random
configuration of atoms, the latter ones show single-atom blocks
which, by construction, only produce those
 background and interference contributions which automatically survive
 the disorder average. In
accordance with this, all arrows in Fig.~\ref{fig:decomp} are for
convenience oriented along the horizontal line.

Prior to presenting the explicit expressions for the
building blocks on the right hand side of Fig.~\ref{fig:decomp}, we
will show that the elastic blocks can be decomposed further.

\subsubsection{Elementary `elastic' building blocks}
\label{sec:irreduc_el} Each of the building blocks in
Fig.~\ref{fig:decomp} contains two outgoing arrows. We will now
express all elastic blocks (circles) as combinations of the blocks
which each contain only one outgoing arrow. We will refer to
such building blocks with one outgoing arrow as `elementary' ones.

\begin{figure}
\includegraphics[width=5cm]{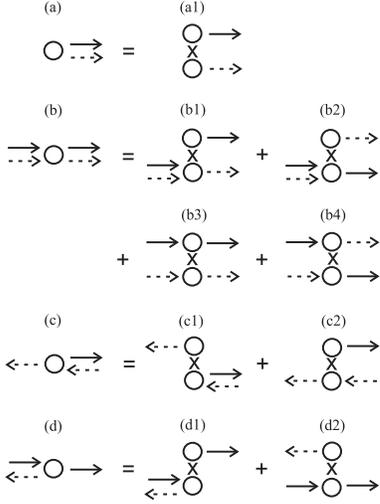}
\caption{Decomposition of the elastic scattering processes into
elementary processes. (a) the elastic intensity of light scattered by a
two-level atom is equal to (a1): the product of the amplitudes for the
elastic scattering of positive and negative frequency amplitudes
described by circles with outgoing solid and dashed arrows; (b) in
case of two incoming probe fields there are four ways, (b1)-(b4),
to distribute the probe fields among the irreducible blocks; (c),
(d) decomposition of the elastic responses in case of one incoming
probe field into (c1), (c2) and (d1), (d2), two elementary blocks, respectively.}
 \label{fig:decomp_el}
\end{figure}

In the case when there are no
incoming arrows, see Fig.~\ref{fig:decomp}(a), the outgoing
arrows correspond to the elastic intensity of light scattered
by the laser-driven atom.
It is well-known \cite{cohen-tannoudji}
that the elastic intensity is proportional to $\la\s^+\s^-\ra^{(0,{\rm el})}=\la\s^+\ra^{(0)}\la\s^-\ra^{(0)}$, that is, given by the product of the expectation values of the atomic dipole operators for an atom driven only by the laser field (we will supply the expectation values in this case by the superscript $(0)$). We represent this equation graphically in Fig.~\ref{fig:decomp_el}(a). There, a circle with the outgoing solid (dashed) arrow corresponds to $\la\s^-\ra^{(0)}$ ($\la\s^+\ra^{(0)}$), respectively, whereas the product of these two amplitudes is depicted
by symbol `$\times$'.

It is easy to generalize this result to the case when there are incoming probe-field amplitudes. In this case, the elastic building blocks
can be expanded into sums of products of two elementary
amplitudes, with the number of terms in this expansion
being equal to the number of ways in which the incoming arrows
can be distributed among the two circles (see
Fig.~\ref{fig:decomp_el}(b-d)).

\subsubsection{`Inelastic' building blocks}
\label{sec:irreduc_in} Concerning the inelastic building blocks of
Fig.~\ref{fig:decomp} (hatched squares),
they result from the non-factorizable part of the atomic response, {\em i.e.},
$\la\s^+\s^-\ra-\la\s^+\ra\la\s^-\ra$, see equation~(\ref{inelcorr}) below. Unlike
the elastic blocks, these blocks, which exhibit two outgoing arrows corresponding to
$\s^+$ and $\s^-$, respectively, cannot be factorized into a product of blocks with only one outgoing arrow.

\subsection{Calculating building blocks}
\label{sec:assessing-single}
In Sec.~\ref{sec:graph-single} we presented the disorder averaged
elastic and inelastic single-atom building blocks which  contribute
to the double scattering background and interference spectra of CBS.
All these blocks represent a single, two-level, laser-driven atom
which additionally receives none, one or two classical
probe-field amplitudes. We will next show how
to evaluate these blocks using the OBE under bichromatic driving.

\subsubsection{Single-atom optical Bloch equations under bichromatic driving}

A classical bichromatic driving field reads 
\begin{equation} {\cal E}(t)={\cal
E}_L^*e^{i\omega_L t}+ {\cal E}_Le^{-i\omega_L
t}+\varepsilon^*e^{i\omega_p t}+\varepsilon e^{-i\omega_p
t}\label{bifield},\end{equation} where ${\cal E}_L$, $\varepsilon$ are complex amplitudes, and $\omega_L$,
$\omega_p$ the frequencies of the pump and probe field,
respectively. It is easy to show that the optical Bloch vector
$\la\vec{\sigma}\ra=(\la\sigma^-\ra,\la\sigma^+\ra,\la\sigma^z\ra)$,
where $\sigma^-=|0\ra\la 1|$, $\sigma^+=|1\ra\la 0|$,
$\sigma^z=|1\ra\la 1|-|0\ra\la 0|$, and $|0\ra$, $|1\ra$ the
ground and excited states of the atom, obeys the following equation
of motion written in the frame rotating at the laser frequency: \begin{equation}
\la\dot{\vec{\sigma}}\ra=M\la\vec{\sigma}\ra+\vec{L}+v\;e^{i\omega
t}\Delta^{(+)} \la\vec{\sigma}\ra+v^*e^{-i\omega
t}\Delta^{(-)}\la\vec{\sigma}\ra,\label{obe}\end{equation} where $M$ denotes the
optical Bloch matrix for an atom driven by the laser field: \begin{equation}
M=\lt(\begin{array}{ccc}-\gamma+i\delta&0&-i\Omega/2\\
0&-\gamma-i\delta&i\Omega^*/2\\-i\Omega^*&i\Omega&-2\gamma\end{array}\rt)\label{eq:BlochM},\end{equation}
with $\delta=\omega_L-\omega_0$ the laser detuning from the atomic
transition frequency $\omega_0$, and $\gamma$ half the spontaneous
decay rate. $\Omega=2{\cal E}_L/\hbar$ and $v=2\varepsilon d/\hbar$
are the Rabi frequencies of the pump and probe fields, respectively,
with $d$ the (real) matrix element of the atomic dipole transition.
$\omega=\omega_p-\omega_L$ is the detuning between the probe and
pump frequencies. Finally, the matrices \begin{equation}
\Dm=\lt(\begin{array}{ccc}0&0&-i/2\\0&0&0\\0&i&0\end{array}\rt),\;\;
\Dp=\lt(\begin{array}{ccc}0&0&0\\0&0&i/2\\-i&0&0\end{array}\rt)\label{eq:defDmDp} \end{equation}
describe the coupling of the atom to the positive and negative frequency
components of the scattered field, respectively, and \begin{equation}
\vec{L}=(0,0,-2\gamma)^T \end{equation} is a constant vector.

\subsubsection{Elastic building blocks}
\label{sec:graph-elastic}
\begin{figure}
\includegraphics[width=8cm]{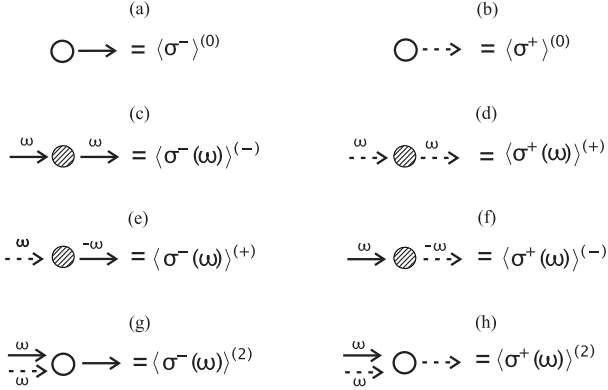}
\caption{Graphical representation of the elementary diagrams for
elastic scattering, and their mathematical expressions in terms of
solutions of equation~(\ref{obe}) for the optical Bloch vector. Waves at the laser frequency
are not labelled. Left and
right columns describe positive- and negative-frequency amplitudes,
respectively, of: (a) and (b) -- elastic scattering in absence of
the probe field; (c) and (d) -- amplitudes of stimulated emission at the probe
field frequency $\omega$ by a laser-driven atom; (e) and (f) --
phase conjugation via nonlinear wave mixing of the pump and
probe fields; (g) and (h) -- amplitudes of stimulated emission at the laser frequency
by an atom subjected to a positive- and negative-frequency probe field
 at the frequency $\omega$.}
\label{fig:zeroth_el}
\end{figure}

It is straightforward to establish the correspondence between the elementary elastic building blocks of Fig.~\ref{fig:decomp_el} and solutions of equation~(\ref{obe}). We will start
by noting that if $v=v^*=0$ in equation~(\ref{obe}), it reduces to the standard OBE having the steady-state solution $\la\vec{\s}\ra^{(0)}$. For non-zero probe fields, there appear corrections to the unperturbed components of the Bloch vector as well as oscillations at the frequencies $\pm \omega$ (see \ref{appA}). Depending on whether a single incoming probe field represents a positive- or negative-frequency amplitude at frequency $\omega$, the corrections will be denoted $\la\vec{\s}(\omega)\ra^{(-)}$ and $\la\vec{\s}(\omega)\ra^{(+)}$, respectively. A second-order correction, which corresponds to one solid and one dashed incoming arrow at the frequency $\omega$, will be denoted as $\la\vec{\s}(\omega)\ra^{(2)}$. Finally, we recall that the emission of the positive-(negative-) frequency amplitudes is associated with the expectation values of the atomic lowering (raising) operators \cite{cohen-tannoudji}. Summarizing the above,
we express the elementary elastic building blocks in terms of the solutions of equation~(\ref{obe}) as shown in Fig.~\ref{fig:zeroth_el}. From now on we use the convention not to label the incoming (outgoing) waves which are exactly on resonance with the laser frequency. Furthermore, we will denote elastic building blocks with outgoing wave at the laser frequency by a blank circle, whereas a hatched circle corresponds to an outgoing wave at a frequency $\omega\neq \omega_L$ (as shown below, the latter requires the presence of an incoming wave with frequency $\omega$ or $-\omega$).

Diagrams (a) and (b) of Fig.~\ref{fig:zeroth_el} describe positive
and negative frequency amplitudes, respectively, of the elastic scattering by
a laser-driven atom. Clearly, their product is nothing but the diagram (a1) in Fig.~\ref{fig:decomp_el}.
 Diagrams (c)-(h) include one
 or two incoming arrows at the frequency $\omega$. Atomic response
 functions depicted by these diagrams were studied by Mollow \cite{mollow72} in the
 context of the weak probe absorption or amplification by a
 laser-pumped two-level atom (see \ref{app:connect}). 
 As follows
 from the analysis presented in \ref{app:connect}, the processes
 (c) and (d) are correspond to the amplitudes of stimulated emission at
the probe field frequency, and hence, their outgoing arrows are labeled by the
same frequency values as the incoming ones. In accord with the convention adopted above, these building blocks are hatched, since the frequency of the incoming wave -- and therefore also of the elastically scattered outgoing wave -- differs from the laser frequency. Further on,
the processes (e) and (f) correspond to a nonlinear phase conjugation whereupon
the positive- (negative-)frequency amplitude at the frequency $\omega$ is scattered into a negative- (positive-)frequency, that is, conjugated, amplitude at the frequency $-\omega$, via a nonlinear mixing with the laser wave described by the nonlinear atomic susceptibility. Using the nonlinear optics terminology, one can regard $\la\s^-(\omega)\ra^{(-)}$,
$\la\s^+(\omega)\ra^{(-)}$ (and their complex conjugate quantities) as, respectively, the effective (that is, dependent on the laser pump) linear and nonlinear susceptibilities of a two-level system with respect to the probe field \cite{boyd}. We note that the relevance of nonlinear susceptibilities for CBS from saturated atoms was mentioned before \cite{gremaud06}.
Finally, the processes (g) and
(h) describe stimulated emission of the amplitudes at the pump field frequency in the presence of the probe field.

We conclude this part by a short remark. The diagrams of Fig.~\ref{fig:zeroth_el} describe all possible elementary building blocks that appear in Fig.~\ref{fig:decomp_el}. Consequently, solutions of equation~(\ref{obe}) for the Bloch vector to second order in the probe-field amplitude allow us to evaluate  completely the `elastic' responses on the right hand side of Fig.~\ref{fig:decomp}.

\subsubsection{Inelastic building blocks}
\label{sec:graph-inelastic} Let us now address the inelastic building blocks. They are
deduced from the solutions of equation~(\ref{obe}) for the atomic dipole
temporal correlation function \cite{geiger10,wellens10} (see also
\ref{appA}), \begin{equation}
g(t_1,t_2)=\la\s^+(t_1)\s^-(t_2)\ra-\la\s^+(t_1)\ra\la\s^-(t_2)\ra,
\label{inelcorr} \end{equation} in the steady-state limit when
$g(t_1,t_2)=g(t_1-t_2)$, and subsequent Laplace transformation to the frequency domain.
\begin{figure}
\includegraphics[width=8cm]{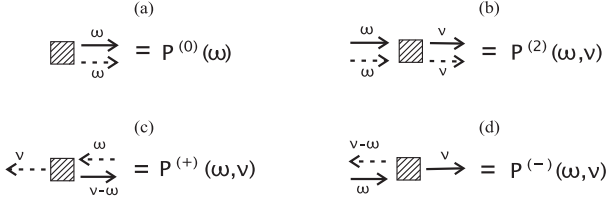}
\caption{Graphical representation of the inelastic building blocks
and their mathematical expressions (see equation~(\ref{eq:inel_func})).
(a) $P^{(0)}(\omega)$ -- the Mollow triplet; (b)
$P^{(2)}(\omega,\nu)$ -- correction to
 the Mollow triplet due to the probe field at the frequency $\omega$; (c)
 $P^{(+)}(\omega,\nu)$ and (d) $P^{(-)}(\omega,\nu)$ -- inelastic wave mixing
 of the pump and probe fields resulting in scattering of the waves neither of which is in resonance with the laser or probe waves.} \label{fig:inel}
\end{figure}

The Laplace transformed solutions of equation~(\ref{inelcorr}), with the probe field amplitudes $v$ and $v^*$ set equal to zero, together with the linear ($\propto v, v^*$) and quadratic ($\propto |v|^2$) corrections, yield the inelastic building blocks shown in Fig.~\ref{fig:inel}.
The physical interpretation of the functions $P^{(0)}(\omega)$, $P^{(\pm)}(\omega,\nu)$ and $P^{(2)}(\omega,\nu)$ is analogous to the interpretation of the building blocks of Fig.~\ref{fig:zeroth_el}, and merely generalizes it to include the multi-photon
scattering of the laser photons.

Diagram (a) in Fig.~\ref{fig:inel} represents the inelastic spectrum of a two-level
atom subject to monochromatic driving. That is, the function
$P^{(0)}(\omega)$ is the
resonance fluorescence spectrum known, for $\Omega\gg \gamma$, as the Mollow triplet
\cite{mollow69}. Diagram (b) represents the correction to the Mollow triplet $P^{(2)}(\omega,\nu)$ proportional to the intensity of the probe field at the
frequency $\omega$.
This diagram can be regarded as a generalization of diagrams
 (g) and (h) of Fig.~\ref{fig:zeroth_el}, which accounts for the inelastic scattering.
 Finally, the functions $P^{(\pm)}(\omega,\nu)$ (see Fig.~\ref{fig:inel}(c,d)) represent inelastic wave mixing processes resulting in the emission of two amplitudes which are off-resonant with either the laser or the probe fields. These diagrams should be contrasted with the diagrams (c,d) in Fig.~\ref{fig:decomp_el}. Indeed, in the latter diagrams, the two outgoing arrows are on-resonant with the probe and laser fields.

\section{Combining building blocks}
\label{sec:connect}
\subsection{Rules}
\label{sec:rules}
In the previous Section we defined all
single-atom building blocks. We recall that the elastic and
inelastic building blocks from which the background and interference
contributions are constructed are shown on the right hand sides of
Fig.~\ref{fig:decomp}(a) and (b), respectively. The four elastic
blocks in Fig.~\ref{fig:decomp} can be further decomposed according
to Fig.~\ref{fig:decomp_el}, with the elementary diagrams defined
in Fig.~\ref{fig:zeroth_el}, whereas the inelastic blocks and their
corresponding spectral response functions are shown in
Fig.~\ref{fig:inel}. So, all diagrams on the right hand side of Fig.~\ref{fig:decomp} are
defined, and we need to introduce rules for combining them.

These rules can be formulated as follows:
\begin{itemize}

\item The ladder and crossed contributions are obtained
by combining the building blocks of atom 1 with that of atom 2, on
the right hand side of Fig.~\ref{fig:decomp}(a) and (b); each
combined two-atom diagram is proportional to the factor
$g^2$, where $g=(k_L\ell)^{-1}$ scales as
the far-field dipole-dipole coupling between the two atoms separated
by the mean distance $\ell$.

\item The elastic building blocks for atoms 1 and 2 are expanded
according to Fig.~\ref{fig:decomp_el}.

\item Each outgoing solid (dashed) arrow of atom 1 is merged with each incoming solid (dashed) arrow of atom
2. As a result, one obtains sums of diagrams containing two
intermediate and two outgoing arrows. The intermediate arrows' frequencies are defined self-consistently: the frequencies of the incoming and outgoing arrows must coincide
and, moreover, respect the form of the response functions presented in Figs.~\ref{fig:zeroth_el}, \ref{fig:inel}. As a result, for example, there is only one independent inelastic frequency describing intermediate amplitudes.

\item The response functions associated with the single atom building
blocks in the resulting double scattering diagrams must be multiplied. The symbolic expressions for these functions are
given in Figs.~\ref{fig:zeroth_el} and \ref{fig:inel}.

\item If an intermediate arrow representing the \emph{inelastic}
probe field (say, at frequency $\omega$) changes its
frequency then the expression corresponding to that diagram
is integrated over this frequency as $\int_{-\infty}^{\infty}d\omega/(2\pi)$ (note that $\omega=\omega_p-\omega_L$, and can take on negative values).

\item Diagrams containing blocks connected by intermediate
counter-propagating arrows with no outgoing arrows must be excluded.
Such diagrams correspond to a photon cycling between the atoms
without contributing to the detected signal. Exactly the same rule exists
in the nonlinear diagrammatic theory of CBS in a
Kerr medium \cite{wellens08}.

\end{itemize}

\begin{figure}
\includegraphics[width=7cm]{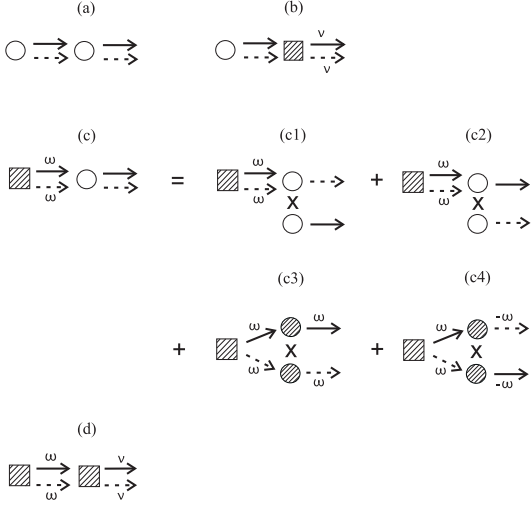}
\caption{Background contributions to double scattering. (a) Fully
elastic contribution originating from both atoms scattering
elastically; (b) inelastic contribution due to second atom
scattering inelastically; (c) mixed contribution including
diagrams (c1), (c2) corresponding to elastic scattering,
and (c3), (c4) corresponding to inelastic scattering;
(d) fully inelastic contribution.}
\label{fig:ladder}
\end{figure}

Using the above rules, we can represent the background and
the interference contributions as sums of the diagrams shown in
Figs.~\ref{fig:ladder} and \ref{fig:crossed}, respectively. When
writing down the explicit expressions for different contributions,
we will omit the overall factor $g^2$.
\subsection{Background contributions}
\subsubsection{Elastic spectrum}

By definition, the elastic background spectrum $L_{\rm el}(\nu)=L_{\rm el}\delta(\nu)$ exhibits a single $\delta$-peak precisely at the laser frequency, corresponding to
$\nu=0$ in the rotating frame.
Hence, it is given by the sum of all
diagrams with unlabeled signal (outgoing) lines in
Fig.~\ref{fig:ladder}, that is, diagrams (a), (c1), and (c2). Note
that diagram (a) is to be unfolded into a sum of products of the
following diagrams of Fig.~\ref{fig:decomp_el}: (a1)(b1), (a1)(b2),
(a1)(b3), and (a1)(b4). Taking the expressions for the building blocks
from Figs.~\ref{fig:zeroth_el}, \ref{fig:inel}, and applying the
above rules for combining these building blocks, the elastic background
intensity reads: \begin{subequations}
\begin{align}
L_{\rm el}&=\la\s^+\ra^{(0)}\la\s^-\ra^{(0)}\la\s^-\ra^{(0)}\la\s^+(0)\ra^{(2)}\n\\
&+\la\s^+\ra^{(0)}\la\s^-\ra^{(0)}\la\s^+\ra^{(0)}\la\s^-(0)\ra^{(2)}\n\\
&+\la\s^+\ra^{(0)}\la\s^-\ra^{(0)}\la\s^+(0)\ra^{(+)}\la\s^-(0)\ra^{(-)}\n\\
&+\la\s^+\ra^{(0)}\la\s^-\ra^{(0)}\la\s^-(0)\ra^{(+)}\la\s^+(0)\ra^{(-)}\label{eq:lel-a}\\
&+\int_{-\infty}^{\infty}\frac{d\omega}{2\pi}P^{(0)}(\omega)\la\s^+(\omega)\ra^{(2)}\la\s^-\ra^{(0)}\label{eq:lelc1}\\
&+\int_{-\infty}^{\infty}\frac{d\omega}{2\pi}P^{(0)}(\omega)\la\s^-(\omega)\ra^{(2)}\la\s^+\ra^{(0)},
\label{eq:lelc2}
\end{align}
\label{ladder-elastic} \end{subequations}
where equations~(\ref{eq:lel-a}),
(\ref{eq:lelc1}) and (\ref{eq:lelc2}) correspond to diagrams (a), (c1) and (c2) in
Fig.~\ref{fig:ladder}, respectively (see \ref{appA} for the explicit evaluation of the right hand side of equation~(\ref{ladder-elastic})).

\subsubsection{Inelastic spectrum}
As evident from Fig.~\ref{fig:ladder}, its remaining diagrams
describe the inelastic spectrum \begin{subequations} \begin{align}
L_{\rm inel}(\nu)&=\la\s^+\ra^{(0)}\la\s^-\ra^{(0)}P^{(2)}(0,\nu)\label{eq:lin-b}\\
&+P^{(0)}(\nu)\la\s^-(\nu)\ra^{(-)}\la\s^+(\nu)\ra^{(+)}\label{eq:lin-c3}\\
&+P^{(0)}(\nu)\la\s^+(\nu)\ra^{(-)}\la\s^-(\nu)\ra^{(+)}\label{eq:lin-c4}\\
&+\int_{-\infty}^{\infty}\frac{d\omega}{2\pi}P^{(0)}(\omega)P^{(2)}(\omega,\nu),\label{eq:lin-d}\end{align}
\label{eq:inel}\end{subequations} where equations~(\ref{eq:lin-b}), (\ref{eq:lin-c3}),
(\ref{eq:lin-c4}), and (\ref{eq:lin-d}) correspond to diagrams (b),
(c3), (c4) and (d) of Fig.~\ref{fig:ladder}, and the (signal)
frequency $\omega$ was for convenience relabeled in lines 2 and 3 of equation~(\ref{eq:inel})
 to $\nu$.
\begin{figure}
\includegraphics[width=8cm]{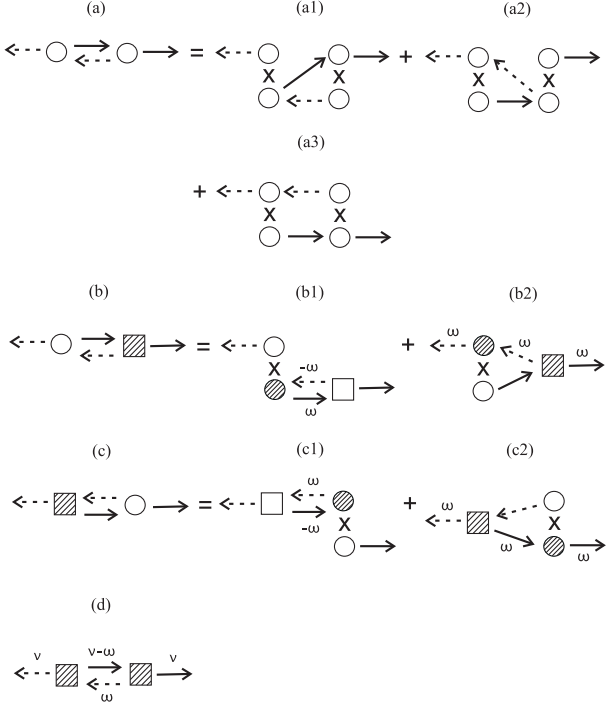}
\caption{Interference contributions to double scattering. (a) Fully
elastic contribution; (b) and (c) mixed contributions including
diagrams (b1), (c1) corresponding to elastic scattering, and
(b2), (c2) corresponding to inelastic scattering; (d) fully inelastic
contribution.} \label{fig:crossed}
\end{figure}

\subsection{Interference contributions}
\subsubsection{Elastic spectrum}
By the same argument as in the case of the background diagrams, we
identify the diagrams (a), (b1), and (c1) of Fig.~\ref{fig:crossed} as
contributing to the elastic spectrum. In the present case, the fully
elastic contribution represented by diagram (a) is explicitly
decomposed into diagrams (a1), (a2), and (a3) corresponding to a
combination of diagrams (c1)(d2), (c2)(d1), and (c2)(d2) of
Fig.~\ref{fig:decomp_el}, respectively. The product  of diagrams
Fig.~\ref{fig:decomp_el}(c1) and (d1) does not contribute to
Fig.~\ref{fig:crossed}(a) because it corresponds to a photon cycling between
the atoms (see Sec.~\ref{sec:rules}).

Note that the squares which are hatched in
Fig.~\ref{fig:crossed}(b), (c) become blank in diagrams (b1) and
(c1). This is precisely the consequence of the self-consistent combination
of diagrams:
 the connection of the phase conjugation diagram Fig.~\ref{fig:zeroth_el}(e), for example, to Fig.~\ref{fig:inel}(d) enforces the occurrence of two frequencies $\omega$ and $-\omega$, and therefore $\nu=0$, in the building block Fig.~\ref{fig:inel}(d). Thus,
 a blank square denotes an inelastic process, where the frequency of one of the outgoing arrows equals the laser frequency.

After these remarks, we present the explicit expression for the
elastic interference spectrum: \begin{subequations} \begin{align} C_{\rm
el}&=\la\s^+\ra^{(0)}\la\s^+\ra^{(0)}\la\s^-(0)\ra^{(+)}\la\s^-(0)\ra^{(-)}\label{eq:cel-a1}\\
&+\la\s^-\ra^{(0)}\la\s^-\ra^{(0)}\la\s^+(0)\ra^{(-)}\la\s^+(0)\ra^{(+)}\label{eq:cel-a2}\\
&+\la\s^-\ra^{(0)}\la\s^+\ra^{(0)}\la\s^-(0)\ra^{(-)}\la\s^+(0)\ra^{(+)}\label{eq:cel-a3}\\
&+\int_{-\infty}^{\infty}\frac{d\omega}{2\pi}\la\s^+\ra^{(0)}\la\s^-(-\omega)\ra^{(+)}
P^{(-)}(\omega,0)\label{eq:cel:b1}\\
&+\int_{-\infty}^{\infty}\frac{d\omega}{2\pi}\la\s^-\ra^{(0)}\la\s^+(-\omega)\ra^{(-)}
P^{(+)}(\omega,0)\label{eq:cel:c1},
\end{align}\label{eq:cr-el}\end{subequations}
where the subsequent lines of equation~(\ref{eq:cr-el}) correspond to
diagrams (a1), (a2), (a3), (b1), and (c1) of Fig.~\ref{fig:crossed},
respectively.

\subsubsection{Inelastic spectrum}
The three contributions to the inelastic spectrum are depicted by
diagrams (b2), (c2) and (d) in Fig.~\ref{fig:crossed}. In diagram
(d), both the solid and dashed lines corresponding to intermediate
amplitudes that change their frequencies. To write down the expression
for this diagram, we note that the outgoing inelastic photon
frequency is $\nu$. Since the frequencies of the
intermediate amplitudes are correlated,
 {\em e.g.}, $\omega$ and $\nu-\omega$ in Fig.~\ref{fig:crossed}(d), there is only one
independent probe field frequency $\omega$ to be integrated over
at fixed detected frequency $\nu$.
Denoting the signal frequency $\nu$ in diagrams (b2) and (c2) of Fig.~\ref{fig:crossed},
finally we
 obtain the following result for the inelastic interference spectrum: \begin{subequations}
\begin{align}
C_{\rm
inel}(\nu)&=\la\s^-\ra^{(0)}\la\s^+(\nu)\ra^{(+)}P^{(-)}(0,\nu)\label{eq:crin-b2}\\
&+\la\s^+\ra^{(0)}\la\s^-(\nu)\ra^{(-)}P^{(+)}(0,\nu)\label{eq:crin-b3}\\
&+\int_{-\infty}^{\infty}\frac{d\omega}{2\pi}P^{(+)}(\omega,\nu)P^{(-)}(\nu-\omega,\nu),\label{eq:crin:d}
\end{align}
\end{subequations} where equations~(\ref{eq:crin-b2}), (\ref{eq:crin-b3}), and
(\ref{eq:crin:d}) correspond to diagrams (b2), (c2), and (d) in
Fig.~\ref{fig:crossed}.

\section{Conclusion}
\label{summary} We presented a diagrammatic derivation of the double
scattering background and interference spectral distributions of
coherent backscattering from two two-level atoms. Although these
distributions were known before \cite{geiger09,wellens10,geiger10}, here we
re-derived them in an intuitive way which, we believe, makes the
diagrammatic approach to CBS of intense laser light from cold atoms
more accessible and attractive to the reader.

It is not only the diagrams that render our present approach
intuitive, but also a physical interpretation of the building blocks
which is given in this work. In particular, we established a connection
of the `elastic' building blocks with the response functions used to calculate the
weak probe absorption spectra by a laser-driven atom
\cite{mollow72}. Furthermore, the `inelastic' building blocks include
the Mollow triplet, a modification thereof proportional to the
intensity of the weak probe field, and inelastic response functions which are linear in the probe fields, and which emerge due to mixing of the
laser and probe waves.

Among the single-atom spectral response functions relevant to CBS, only the resonance fluorescence and probe absorption spectra were observed experimentally \cite{schuda74,wu77}. It remains to be seen whether the other functions can also be measured directly in experiments with single quantum scatterers interacting with one strong and one weak laser field. Perhaps, strongly driven quantum dots
\cite{xu2007} are most suitable candidates for such observations.

The possibility of obtaining the double scattering signal diagrammatically from single-atom building blocks suggests two directions of future research, in the field of CBS of intense laser light from cold atoms. First, the diagrammatic techniques presented in this work will be further developed to assess higher scattering orders of CBS. Second, a generalization of this method to realistic atoms having degenerate dipole transitions will be another useful application.

\ack
Partial financial support by the DFG Research Grant 760 is gratefully acknowledged. V.S. acknowledges financial support through DFG grant BU-1337/9-1.

\appendix

\section{Expressions for building blocks}
\label{appA}
Here we present expressions for the elementary elastic and inelastic building blocks (Figs.~\ref{fig:zeroth_el} and \ref{fig:inel}, respectively). A detailed derivation thereof can be found, for example, in \cite{geiger10}. 
\subsection{Elastic blocks}
\label{app:elastic}
The expressions associated with the building blocks of Fig.~\ref{fig:zeroth_el} follow from the perturbative (in the probe field amplitude) solutions of the optical Bloch equation (\ref{obe}). They are given by the first or second entries of the following vectors:
\begin{subequations}
\begin{align}
\la\vec{\s}\ra^{(0)}&=G\vec{L},\\
\la\vec{\s}(\omega)\ra^{(+)}&=G(i\omega)\Dp \la\vec{\s}\ra^{(0)},\\
\la\vec{\s}(\omega)\ra^{(-)}&=G(-i\omega)\Dm \la\vec{\s}\ra^{(0)},\\
\la\vec{\s}(\omega)\ra^{(2)}&=G\Dp \la\vec{\s}(\omega)\ra^{(-)}+ G\Dm \la\vec{\s}(\omega)\ra^{(+)},
\end{align}
\label{sigma012} \end{subequations} where $G(x)$ is the free propagator given by
\begin{equation} G(x)=(x-M)^{-1},\end{equation} where $M$ is the Bloch matrix (see
equation~(\ref{eq:BlochM})), $G\equiv G(0)$, and the matrices $\Delta^{(\pm)}$ describe the coupling of the atomic dipole to the weak probe fields (see equation~(\ref{eq:defDmDp})). For the Bloch vector
$\la\vec{\s}\ra=(\la\s^-\ra,\la\s^+\ra,\la\s^z\ra)$, the
zeroth-order elastic scattering amplitude $\la\s^-\ra^{(0)}$ is
given by $(G\vec{L})_1$. Likewise, using (\ref{sigma012}), one finds
all of the amplitudes of Fig.~\ref{fig:zeroth_el}.

\subsection{Inelastic blocks}
\label{app:inelastic}
The expressions for the four elementary inelastic building blocks of Fig.~\ref{fig:inel} can be found by solving the equations of motions for the two-time correlation vectors:
\begin{subequations}
\begin{align}
\vec{g}_1(t_1,t_2)&=\la\vec{\s}(t_1)\s^-(t_2)\ra-\la\vec{\s}(t_1)\ra\la\s^-(t_2)\ra,\; t_1>t_2,\\
\vec{g}_2(t_1,t_2)&=\la\s^+(t_1)\vec{\s}(t_2)\ra-\la\s^+(t_1)\ra\la\vec{\s}(t_2)\ra,\; t_1<t_2.
\end{align}
\label{2timecorr}
\end{subequations}
By virtue of the quantum regression theorem the vectors $\vec{g}_i(t_1,t_2)$ of equation~(\ref{2timecorr}) satisfy equations of motion similar to the OBE (\ref{obe}), which can be solved by Laplace transformation. The resulting expressions for the inelastic building blocks are given as a sum of two terms
corresponding to equation~(\ref{2timecorr}a) and (\ref{2timecorr}b), respectively:
\begin{subequations}
\begin{align}
P^{(0)}(\omega)&=[\vec{\tilde{g}}^{(0)}_2(-i\omega)]_1+[\vec{\tilde{g}}^{(0)}_1(i\omega)]_2,\\
P^{(-)}(\omega,\nu)&=[\vec{\tilde{g}}^{(-)}_2(\omega,i\omega-i\nu)]_1+[\vec{\tilde{g}}^{(-)}_1(\omega,i\nu)]_2,\\
P^{(+)}(\omega,\nu)&=[\vec{\tilde{g}}^{(+)}_2(\omega,-i\nu)]_1+[\vec{\tilde{g}}^{(+)}_1(\omega,i\nu-i\omega)]_2,\\
P^{(2)}(\omega,\nu)&=[\vec{\tilde{g}}^{(2)}_2(\omega,-i\nu)]_1+[\vec{\tilde{g}}^{(2)}_1(\omega,i\nu)]_2,
\end{align}
\label{eq:inel_func}\end{subequations} where the Laplace transform solutions read:
\begin{subequations}
\begin{align}
\vec{\tilde{g}}_{1,2}(z)^{(0)}&=G(z)\vec{g}^{(0)}_{1,2}(0),\\
\vec{\tilde{g}}_{1,2}(\omega,z)^{(+)}&=G(z+i\omega)[\Dp \vec{\tilde{g}}^{(0)}_{1,2}(z)+\vec{g}^{(+)}_{1,2}(\omega,0)],\\
\vec{\tilde{g}}_{1,2}(\omega,z)^{(-)}&=G(z-i\omega)[\Dm \vec{\tilde{g}}^{(0)}_{1,2}(z)+\vec{g}^{(-)}_{1,2}(\omega,0)],\\
\vec{\tilde{g}}_{1,2}(\omega,z)^{(2)}&=G(z)[\Dp \vec{\tilde{g}}^{(-)}_{1,2}(\omega,z)+\Dm \vec{\tilde{g}}^{(+)}_{1,2}(\omega,z)\n\\
&+\vec{g}^{(2)}_{1,2}(\omega,0)].
\end{align}
\label{lapsol}
\end{subequations}
The initial conditions in the right hand side of equation~(\ref{lapsol}) result from setting $t_1=t_2$ in (\ref{2timecorr}):
\begin{subequations}
\begin{align}
\vec{g}^{(0)}_{1}(0)&=-i\Dp \la\vec{\s}\ra^{(0)}+\vec{L}_1-\la\vec{\s}\ra^{(0)}\la\s^-\ra^{(0)},\\
\vec{g}^{(0)}_{2}(0)&=i\Dm \la\vec{\s}\ra^{(0)}+\vec{L}_2-\la\vec{\s}\ra^{(0)}\la\s^+\ra^{(0)},\\
\vec{g}_1^{(\pm)}(\omega,0)&=-i\Dp\la\vec{\s}(\omega)\ra^{(\pm)}-\la\vec{\s}(\omega)\ra^{(\pm)}\la\s^-\ra^{(0)}\n\\
&-\la\vec{\s}\ra^{(0)}\la\s^-(\omega)\ra^{(\pm)},\\
\vec{g}_2^{(\pm)}(\omega,0)&=i\Dm\la\vec{\s}(\omega)\ra^{(\pm)}-\la\vec{\s}(\omega)\ra^{(\pm)}\la\s^+\ra^{(0)}\n\\
&-\la\vec{\s}\ra^{(0)}\la\s^+(\omega)\ra^{(\pm)},\\
\vec{g}_1^{(2)}(\omega,0)&=-i\Dp\la\vec{\s}(\omega)\ra^{(2)}-\la\vec{\s}(\omega)\ra^{(2)}\la\s^-\ra^{(0)}\n\\
&-\la\vec{\s}\ra^{(0)}\la\s^-(\omega)\ra^{(2)}-\la\vec{\s}(\omega)\ra^{(+)}\la\s^-(\omega)\ra^{(-)}\n\\
&-\la\vec{\s}(\omega)\ra^{(-)}\la\s^-(\omega)\ra^{(+)},\\
\vec{g}_2^{(2)}(\omega,0)&=i\Dm\la\vec{\s}(\omega)\ra^{(2)}-\la\vec{\s}(\omega)\ra^{(2)}\la\s^+\ra^{(0)}\n\\&-\la\vec{\s}\ra^{(0)}\la\s^+(\omega)\ra^{(2)}-\la\vec{\s}(\omega)\ra^{(-)}\la\s^+(\omega)\ra^{(+)}\n\\
&-\la\vec{\s}(\omega)\ra^{(+)}\la\s^+(\omega)\ra^{(-)},
\end{align}
\end{subequations}
with $\vec{L}_1=(0,1/2,0)^T$, and $\vec{L}_2=(1/2,0,0)^T$.

\section{Relation of the building blocks to the results of Mollow}
\label{app:connect}
Apart from notation, the elastic building blocks are equivalent to the results obtained by Mollow, in connection with the study of stimulated emission and absorption of a weak probe field by a coherently pumped two-level atom \cite{mollow72}. Since Mollow's derivation and definitions differ from ours, we here establish the relation between his and our results. Upon the replacements 
\beq
\kappa&\rightarrow& 2\gamma, \; \kappa^\prime\rightarrow \gamma,\; \Delta\omega\rightarrow\Delta,\;   \n\\
\Delta\nu&\rightarrow& \omega, \;2|\lambda\mathcal{ E}_0|\rightarrow\Omega,\; \lambda^\prime\mathcal{ E}^\prime_0\rightarrow v,\n
\eq     
where the left and right hand sides of each substitution correspond to the results of \cite{mollow72} and of our present work, respectively,
we obtain the following identities:
\begin{align}
\bar{\alpha}&\equiv \la\s^-\ra^{(0)}, & (\bar{\alpha})^*&\equiv \la\s^+\ra^{(0)},\label{a1}\\
\delta\alpha_+&\equiv v\la\s^-(\omega)\ra^{(-)}, &(\delta\alpha_+)^*&\equiv v^*\la\s^+(\omega)\ra^{(+)},\label{eq:stimprobe}\\
\delta\alpha_-&\equiv v\la\s^+(\omega)\ra^{(-)},& 
(\delta\alpha_-)^*&\equiv v^*\la\s^-(\omega)\ra^{(+)},\\
\delta\alpha_0&\equiv 4v^2\la\s^-(\omega)\ra^{(2)}, &(\delta\alpha_0)^*&\equiv 4(v^*)^2\la\s^+(\omega)\ra^{(2)}.\label{eq:stimpump}\end{align}
We note that the right hand side of each identity is equal, up to a prefactor, to the elementary building block from Fig.~\ref{fig:zeroth_el}. For completeness, we also provide the expressions
on which the elementary building blocks depend implicitly:
\begin{align}
\bar{n}-\bar{m}&\equiv\la\s^z\ra^{(0)},& \eta&\equiv 2v\la\s^z(\omega)\ra^{(-)},\\
\delta \bar{n}&\equiv 2 |v|^2\la\s^z(\omega)\ra^{(2)},
\end{align}
where, as in equations (\ref{a1})-(\ref{eq:stimpump}), the left and right hand sides of each identity refer to our present results and those of \cite{mollow72}, respectively. 

As shown in \cite{mollow72}, the rate $\mathcal{ W}^{\prime}$ of absorption/stimulated emission of the probe field  is given by
\begin{equation}
\mathcal{ W}^{\prime}=-i\lambda^\prime\mathcal{ E}_0^{\prime*}\delta\alpha_++
i\lambda^\prime\mathcal{ E}_0^{\prime}(\delta\alpha_+)^*.
\end{equation}
Hence, the identities (\ref{eq:stimprobe}) imply our interpretation of the building blocks in Fig.~\ref{fig:zeroth_el}(c,d) as describing a stimulated emission of the probe field in presence of the pump field (whether the field is actually emitted or absorbed is defined by the sign of $\mathcal{ W}^{\prime}$ \cite{mollow72}). 
Analogously, the rate $\delta\mathcal{ W}$ of absorption/stimulated emission of the pump field reads
\begin{equation}
\delta\mathcal{ W}=-i\lambda^* \mathcal{E}_0^*\delta\alpha_0+i\lambda\mathcal{E}_0
(\delta\alpha_0)^*.
\end{equation}  
Therefore, due to the identities (\ref{eq:stimpump}), we interpret the amplitudes $\la\s^\pm(\omega)\ra^{(2)}$ (see Fig.~\ref{fig:zeroth_el}(g,h)) as describing stimulated emission of the pump field in presence of the probe field. 

The physical interpretation of the function $\delta\alpha_-$ ($(\delta\alpha_-)^*$) was not considered by Mollow. But, by analogy with the above results, and using the fact it represents a correction to the negative- (positive-)frequency scattering amplitude for a pumped atom probed by a positive- (negative-)frequency field, we deduce that $\delta\alpha_-$ ($(\delta\alpha_-)^*$) and, consequently, $\la\s^+(\omega)\ra^{(-)}$  ($\la\s^-(\omega)\ra^{(+)}$), see Fig.~\ref{fig:zeroth_el}(e,f), describes a phase-conjugation due to non-linear mixing of the pump and probe fields.


\section*{References}
\begin{thebibliography}{99}

\bibitem{sheng}
Sheng P~1995 {\it Introduction to Wave Scattering, Localization and
Mesoscopic Phenomena} (Academic Press, San Diego).


\bibitem{albada85}
Van Albada M~P and Lagendijk A~1985 \PRL {\bf 55} 2692; Wolf P-E~and 
Maret G 1985 \PRL {\bf 55} 2696;
Kuga Y~and Ishimaru A 1984 \JOSA A {\bf 1} 831.

\bibitem{labeyrie99}
Labeyrie G, de Tomasi F, Bernard J-C, M\"uller C A, Miniatura C
and Kaiser R 1999 \PRL {\bf 83} 5266.




\bibitem{chaneliere04}
Chaneli\`ere T, Wilkowski D, Bidel Y, Kaiser R and Miniatura C 2004
\PR~E {\bf 70} 036602.


\bibitem{balik05}
Balik S, Kulatunga P, Sukenik C~I, Havey M~D, Kupriyanov D~V
and Sokolov I~M~2005 {\it J.~Mod.~Opt.} {\bf 52} 2269.

\bibitem{cohen-tannoudji}
Cohen-Tannoudji C., Dupont-Roc J and Grynberg G 2004 {\it Atom-Photon Interactions}
 (Wiley-VCH, Weinheim).

\bibitem{lehmberg70}
Lehmberg R~H 1970 \PR A {\bf 2} 883.

\bibitem{agarwal_book}
Agarwal G~S~ 1974 {\it Quantum Statistical Theories of
Spontaneous Emission and their Relation to Other Approaches} (Springer, Berlin).

\bibitem{shatokhin05}
Shatokhin V, M\"uller C~A~ and Buchleitner A 2005 \PRL {\bf 94}
043603; Shatokhin V,  M\"uller C~A~ and Buchleitner A 2006 \PR~A {\bf
73} 063813.

\bibitem{shatokhin07}
Shatokhin V, Wellens T,  Gr\'emaud B~and Buchleitner A 2007 \PR A {\bf 76} 043832.

\bibitem{geiger09}
Geiger T 2009 {\it New Approach to Multiple Scattering of Intense Laser Light
from Cold Atoms} Diploma Thesis (Albert-Ludwigs Universit\"at
Freiburg, Freiburg) http://www.freidok.uni-freiburg.de/volltexte/6986

\bibitem{wellens10}
Wellens T, Geiger T, Shatokhin V and Buchleitner A  2010 \PR A {\bf 82}
013832.

\bibitem{geiger10}
Geiger T, Wellens T, Shatokhin V and Buchleitner A  2010
 {\it Photon.
Nanostr. Fund. Appl.} {\bf 8} 244.

\bibitem{shatokhin10}
Shatokhin V, Geiger T, Wellens T  and Buchleitner A  2010 {\it Chem. Phys.} {\bf 375},
150.

\bibitem{mollow72} 
Mollow B~R~1972 \PR A {\bf 5} 2217.

\bibitem{xu2007}
Xu X et al. 2007 {\it Science} {\bf 317} 5840.


\bibitem{kimble77}
Kimble H~J, Dagenais  M and Mandel L 1977 \PRL {\bf 39} 691.

\bibitem{boyd}
Boyd R.~W~2003 {\it Nonlinear Optics} 2nd Ed. (Academic Press, San Diego) Chap. 6.


\bibitem{gremaud06}
Gr\'emaud B,~Wellens T, Delande D and Miniatura C 2006 \PR~A
{\bf 74} 033808.

\bibitem{mollow69}
Mollow B~R~1969 \PR {\bf 188} 1969 .


\bibitem{wellens08} 
Wellens T~and Gr\'emaud B~2008 \PRL {\bf 100} 033902; Wellens T~and Gr\'emaud B  2009 \PR~A {\bf 80} 063827.

\bibitem{schuda74}
Schuda F, Stroud C~R and Hercher M 1974 {\it J. Phys.} B {\bf 7} L198; Wu F~Y, Grove R~E~and Ezekiel S 1975 \PRL {\bf 35} 1426.


\bibitem{wu77}
Wu F~Y, Ezekiel S, Ducloy M and Mollow B~R 1977 \PRL {\bf 38} 1077.




\end{thebibliography}
\end{document}